\documentclass{article}

\usepackage[T1]{fontenc}
\usepackage[utf8]{inputenc}

\usepackage{amsmath, amssymb}
\usepackage{physics}
\usepackage{siunitx}

\usepackage{graphicx}
\usepackage{float}
\usepackage{subcaption}
\usepackage{caption}

\usepackage{hyperref}
\usepackage{doi}
\usepackage{orcidlink}

\usepackage{geometry}
\hypersetup{
 pdftitle={Excitonic Coupling and Photon Antibunching in Venus~Yellow~Fluorescent~Protein Dimers: A~Lindblad~Master~Equation~Approach},
 colorlinks=true,
 linkcolor=blue,
 urlcolor=blue,
 citecolor=blue
}

\title{Excitonic Coupling and Photon Antibunching in Venus~Yellow~Fluorescent~Protein Dimers: A~Lindblad~Master~Equation~Approach}
\author{
Ian T. Abrahams\,\textsuperscript{\orcidlink{0009-0005-7575-3570}} \\
Quantum Biology Doctoral Training Centre \\
University of Surrey, Guildford GU2 7XH, United Kingdom \\
\href{mailto:i.abrahams@surrey.ac.uk}{i.abrahams@surrey.ac.uk}
}
\date{}
\begin{document}
\maketitle

\begin{abstract}
Strong excitonic coupling and photon antibunching (AB) have been observed together in Venus yellow fluorescent protein dimers and currently lack a cohesive theoretical explanation. In 2019, Kim \emph{et al.} demonstrated Davydov splitting in circular dichroism spectra, revealing strong J-like coupling, while antibunched fluorescence emission was confirmed by combined antibunching--fluorescence correlation spectroscopy (AB/FCS fingerprinting). To investigate the implications of this coexistence, Venus yellow fluorescent protein (YFP) dimer population dynamics are modeled within a Lindblad master equation framework, testing its ability to cope with typical, data-informed, Venus YFP dimer time and energy values. Simulations predict multiple-femtosecond (fs) decoherence, yielding bright/dark state mixtures consistent with antibunched fluorescence emission at room temperature. Thus, excitonic coupling and photon AB in Venus YFP dimers are reconciled without invoking long-lived quantum coherence. However, clear violations of several Lindblad approximation validity conditions appear imminent, calling for careful modifications to choices of standard system and bath definitions and parameter values.
\end{abstract}

Like Brownian motion historically, strong excitonic coupling and photon antibunching (AB) in Venus dimers present a set of experimental observations lacking a cohesive theoretical explanation \cite{Einstein1956_Brownian,Kim2019_VenusDimers}. In 2019, Kim \emph{et al.} revealed Davydov splitting in circular dichroism spectra of Venus$_\text{A206}$ (aka dVenus) yellow fluorescent protein (YFP) dimers, measuring strong, negative excitonic coupling between identical chromophore pairs \cite{Kim2019_VenusDimers}, although it was later revealed their choice of monomeric control likely resulted in inflated measured magnitude, compared to a more refined follow-up measurement, including two additional fluorescent protein (FP) dimer variants \cite{Nguyen2025ExcitonicFPs}. Negative coupling can in principle arise from antiparallel, more Venus-like dimer orientations (see Eq. \ref{eq:PDA_J}) \cite{Wang2019}, or conventional head-to-tail geometries \cite{Kasha1963_ExcitonModel,Spano1990_Superradiance,Spano2017_AggregatesBeyondKasha,Hestand2018_HJAggregates,Banerjee2025_DimerCoherence,Davydov1964_Eng}, each corresponding to opposite energy ordering for corresponding sets of bright and dark excitonic states.

Kim \emph{et al.} directly observed photon AB in engineered Venus ``tandem dimers,'' (TD) with enforced hydrophobic dimerization, using a custom antibunching--fluorescence correlation spectroscopy technique (AB/FCS fingerprinting), with evidence of distance dependence of photon emission statistics \cite{Kim2019_VenusDimers}. AB is a hallmark of single-photon emission and, in principle, can arise from strongly coupled dimers collectively absorbing in delocalized eigenstates of the excitonic basis, which are predicted to lose their coupling strength as a function of distance, along with their collective behavior as a consequence \cite{Banerjee2025_DimerCoherence}.

Several theoretical frameworks have previously been applied to molecular dimers \cite{Kasha1963_ExcitonModel,Spano2017_AggregatesBeyondKasha,Hestand2018_HJAggregates,Rouse2019_DarkStates,Banerjee2025_DimerCoherence,Freed2025_PermDipoles}, including recent models of green fluorescent protein (GFP) dimers that emphasize decoherence and bath relaxation \cite{BurgessFlorescu2024_arXiv}. Yet none directly address how strong coupling, spectral shifts, and photon statistics coexist in Venus dimers at room temperature.

Here, I further ask related two main questions: does the inflated $J$ coupling value measurement from the self-contained 2019 Venus study \cite{Kim2019_VenusDimers,Nguyen2025ExcitonicFPs}, combined with experimentally informed values for FP system and bath parameters 1) break the Lindblad (Born--Markov--secular) and high-temperature assumptions, posing fundamental issues to the physical realism of its modeling and/or 2) result in major discrepancies with prior experimental observations? The answer to either of which being yes, hence indicates the need for fundamental changes in parameter choice, and/or any or all of the chosen approximations to be at least partially abandoned in the fluorescent protein (FP) dimer energy and time regimes.

\section{Open Quantum Systems Model}
\label{sec:model}

\subsection{Excitonic Basis and Hamiltonian}

Let $\ket{1}$ and $\ket{2}$ denote localized excitations of two chromophores. In the strong-coupling regime these combine to form delocalized excitonic states,
\begin{equation}
  \ket{+}=\frac{\ket{1}+\ket{2}}{\sqrt{2}}, \qquad
  \ket{-}=\frac{\ket{1}-\ket{2}}{\sqrt{2}},
\end{equation}
corresponding to the bright (symmetric) and dark (antisymmetric) states, respectively. Such coherent superpositions of localized excitations are described in standard treatments \cite{May2011} and reflect a broader perspective \cite{Scholes2003}, emphasizing the central role of delocalization and coherence in molecular excitonic systems. Such delocalization is, however, strongly limited by disorder and thermal fluctuations \cite{Scholes2020}.

The corresponding system Hamiltonian can be defined in the site basis as,
\begin{equation}
\hat{H} = \tfrac{\Delta}{2}\sigma_z + J\sigma_x.
\end{equation}
where $\Delta$ corresponds to the energy difference between the two sites (Appendix \ref{sec:appendix_site_energy_gap_stokes_shift}), and $J$ corresponds to the Coulombic (electronic) coupling energy between the two sites (Appendix \ref{sec:appendix_splitting}), which, following diagonalization, gives eigenenergies,
\begin{equation}
  E_\pm = \pm \tfrac{1}{2}\Delta E,
\end{equation}
where $\Delta E = \sqrt{\Delta^2+4J^2}$ (Appendix \ref{sec:appendix_rotation_site_basis_Ham}). 

For a homodimer ($\Delta=0$), $\Delta E=2|J|$, so that $E_\pm=\pm J$. For negative coupling ($J<0$), the bright state $\ket{+}$ lies lower in energy than the dark state $\ket{-}$.

\subsection{Lindblad Formalism}

The reduced system density operator $\hat{\rho}(t)$ is evolved using a Lindblad master equation in the Born--Markov--secular approximation, as if in the high-temperature limit \cite{Breuer2002}, however, with several underlying assumptions strictly violated (see Appendix \ref{sec:appendix_parameters} for system and bath parameter values and Appendices \ref{sec:appendix_rates} and \ref{sec:appendix_derivations_therm+deph} for additional definitions and derivations),
\begin{equation}
  \frac{d\hat{\rho}}{dt} = -\frac{i}{\hbar}[\hat{H},\hat{\rho}]
  + \sum_{l\in\{\phi,-+,+-\}} \mathcal{L}(\hat{\rho},\gamma_l),
\end{equation}
where 
\[
\gamma_\phi \;=\; \frac{2\,\lambda\,k_B T}{\hbar^2\,\gamma_c}
\]
is the Drude--Lorentz pure-dephasing rate, with $\lambda$ as the effective reorganization energy of the system on the fast timescale, $k_B$ as Boltzmann's constant, $T$ as the temperature of the bath, $\hbar$ as the reduced Planck's constant, and $\gamma_c$ as the cutoff frequency for bath memory. Furthermore, $\gamma_{-+}$ and $\gamma_{+-}$ are the downhill and uphill thermal transfer rates, respectively. These rates are implemented with energy-basis jump operators $L_{-+}=\sqrt{\gamma_{-+}}\,|-\rangle\!\langle+|$ (upper$\to$lower) and $L_{+-}=\sqrt{\gamma_{+-}}\,|+\rangle\!\langle-|$ (lower$\to$upper), with
\[
\gamma_{-+}=S_{\mathrm{cl}}(\omega_0)\big(n_{\mathrm{BE}}(\omega_0,T)+1\big),\quad
\gamma_{+-}=S_{\mathrm{cl}}(\omega_0)\,n_{\mathrm{BE}}(\omega_0,T),
\]
\[
n_{\mathrm{BE}}(\omega_0, T)=\frac{1}{e^{\hbar\omega_0/k_B T}-1},\qquad
S_{\mathrm{cl}}(\omega_0)=\frac{2\,\lambda\,k_B T}{\hbar^2}\,
\frac{\gamma_c}{\gamma_c^2+\omega_0^2},\qquad
\omega_0=\Delta E/\hbar,
\]
where $n_{\mathrm{BE}}(\omega_0,T)$ is the Bose--Einstein occupancy number with transition frequency $\omega_0$ and at temperature $T$, and $S_{\mathrm{cl}}(\omega_0)$ is the high-temperature limit of the bath spectral density. An explicit derivation of the $\gamma_\phi$ dephasing expression, demonstrating its continuity with the thermal transfer rates above, is given in Appendix~\ref{sec:appendix_rates}. Further additional details can be found in Appendix \ref{sec:appendix_derivations_therm+deph}, showing this model as the simplistic limiting case with uncorrelated sites. A full table of simulation parameters can be found in Appendix \ref{sec:appendix_parameters}.

The bath correlation parameters that determine these rates, namely the cutoff frequency $\gamma_c$, and the effective reorganization energy $\lambda$, were extracted from experimental coherence times of a representative GFP reported by Cinelli \emph{et al.}\ \cite{Cinelli2001_PRL} and the dielectric relaxation model of Burgess and Florescu \cite{BurgessFlorescu2024_arXiv} as shown in Appendix \ref{sec:appendix_lambda_fast}. The pure dephasing itself is explicitly derived in Appendix \ref{sec:appendix_dephasing_rate}. It can additionally be noted that when the initial state is diagonal in the energy basis, the thermal jumps defined above preserve diagonality and the dynamics reduce exactly to a two-state Pauli master equation with detailed balance shown explicitly in Appendix \ref{sec:appendix_thermal_transfer}.

\section{Results and Discussion}
\label{sec:results}

Numerical simulations of open quantum systems were performed using QuTiP v5.2.0 \cite{Johansson2012_QuTiP,Johansson2013_QuTiP2}, with supporting scientific computing provided by NumPy v2.3.2 \cite{Harris2020_NumPy} and SciPy v1.16.0 \cite{Virtanen2020_SciPy}. (See Appendix \ref{sec:appendix_time_step_window} for further details on time-step and window selection.)

\subsection{Subpicosecond Dynamics, Photon Antibunching, and Absorption Redshift}
\label{sec:results_antibunching}

\begin{figure}[h!]
  \centering
  \includegraphics[width=1\linewidth]{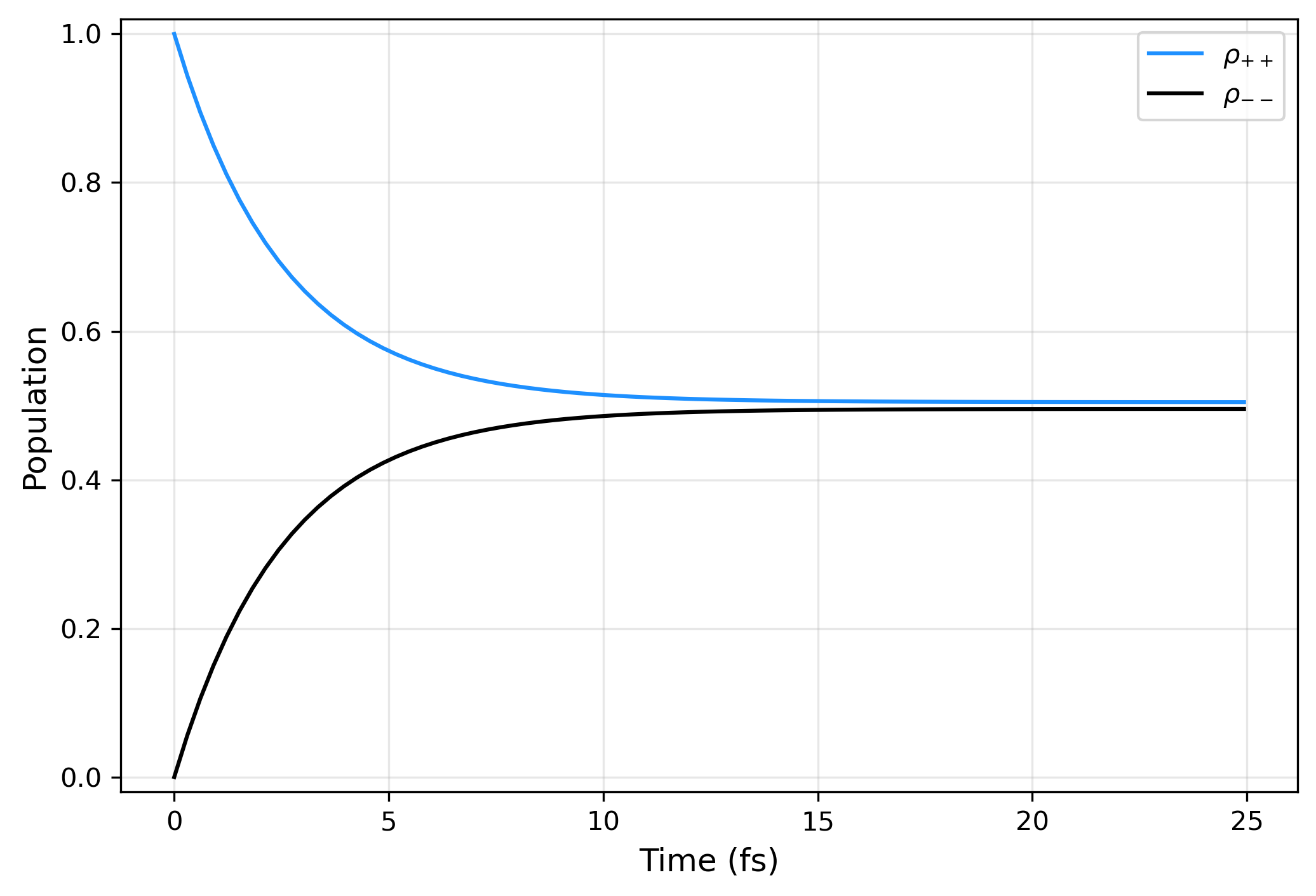}
  \caption{\textbf{Room-temperature, sub-ps thermal relaxation dynamics.} 
Population dynamics of a Venus dimer at room temperature ($T = 293$~K) shown in the energy (exciton) basis, initialized in the bright state ($\rho_{++}$). In this true-homodimer case, $\Delta=0$, representing a timescale prior to vibrational relaxation. The exciton splitting $\Delta E=2|J|$, where the Coulombic coupling energy $J = -34$~meV (Appendix \ref{sec:appendix_splitting}). Inhomogeneous site decoherence occurs in $T_2^*\approx2.77$ fs, resulting in visibly overdamped, monotonic population decay.}
\label{fig:room-temperature_subps_dynamics}
\end{figure}

I now analyze subpicosecond (sub-ps) exciton population dynamics for a Venus dimer model system (see Fig. \ref{fig:room-temperature_subps_dynamics}) and its implications for photon antibunching and absorption redshift observations.

When prepared in a pure bright state with $\Delta=0$, the system equilibrates to a near 50:50 mixed state within 25 fs (see Fig. \ref{fig:room-temperature_subps_dynamics}), following inhomogeneous site decoherence in $T_2^*\approx2.77$ fs, calculated using $|\rho_{12}|\approx e^{1/T_2^*}$, near the fundamental lower bound for the timescale of absorption of visible light \cite{Wolf2017_NatCommun,Kumpulainen2017_ChemRev}, indicating likely nonlocal absorption into bright excitonic eigenstates \cite{SpanoMukamel1989}, rather than individual sites, linking directly to observed photon AB \cite{Kim2019_VenusDimers,Nguyen2025ExcitonicFPs}. Thermal conditions of the protein also likely result in transient site energy differences and slight shifts in dipole orientation causing average early-time excitonic state fluctuations. Such fluctuations plausibly explain the observed reduced absorption redshift of $\sim4$ meV relative to the expected $\sim34$ meV for a true Venus J-dimer, as modeled, while still effectively initialized in a pure bright state. A more accurate model, both structurally and energetically, however, could yield a significantly different prediction of specific dynamics, including the dephasing time $T_2$, potentially even allowing for previously undetected, ultrafast superradiance \cite{Kim2019_VenusDimers,Nguyen2025ExcitonicFPs,Spano1990_Superradiance}.

\subsection{Thermal Dynamics from a Mixed State and Brightness}\label{sec:results_mixed}

To focus on long-timescale population dynamics ($t\gtrsim100$ fs) of the present model, the Venus dimer model system is initialized as a 50:50 diagonal mixed state in the exciton basis, $\rho_0=\tfrac12(|+\rangle\!\langle+|+|-\rangle\!\langle-|)$ , which coarse-grains the sub-ps redistribution after bright-state preparation, focusing on the ps-scale, dominated by thermal transfer dynamics (Appendix \ref{sec:appendix_thermal_transfer}), picking up directly after Figure \ref{fig:room-temperature_subps_dynamics}.

\begin{figure}[h!]
  \centering
  \includegraphics[width=1\linewidth]{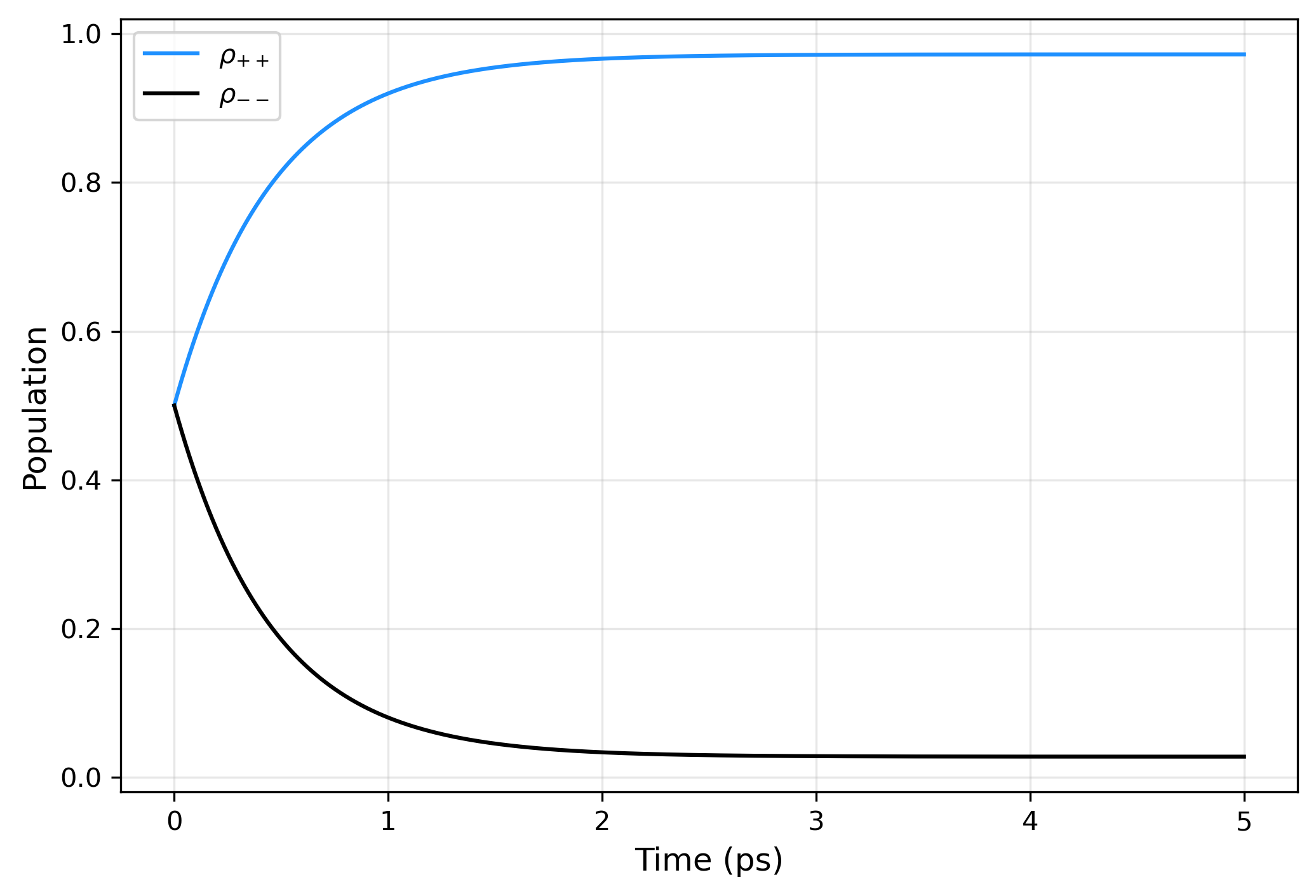}
  \caption{\textbf{Room-temperature, ps thermal relaxation dynamics from an equal bright--dark mixture.} 
Population dynamics of a Venus dimer at room temperature ($T = 293$~K) shown in the energy (exciton) basis, initialized in a 50:50 bright--dark mixed state ($\rho_{++}=\rho_{--}=0.5$), excluding pure dephasing. In this effective-heterodimer case,  $\Delta = 59\ \mathrm{meV}$, representing the Venus Stokes shift following vibrational relaxation (Appendix \ref{sec:appendix_site_energy_gap_stokes_shift}). The exciton splitting becomes $\Delta E=\sqrt{\Delta^2+4J^2} \approx 90$ meV, where the Coulombic coupling energy  $J = -34$~meV, favoring the bright exciton state, although, for a mixed state, both bright and dark exciton states should be optically accessible.}
\label{fig:energy_mixed_state_dynamics}
\end{figure}
After sub-ps relaxation, initializing a 50:50 bright--dark mixture yields a strong thermal bias toward the bright state (see Fig. \ref{fig:energy_mixed_state_dynamics}). The corresponding exciton splitting and eigenenergies are given in Sec.~\ref{sec:model}, now with $\Delta \neq 0$ determined by the characteristic Stokes shift, following protein vibrational relaxation (Appendix \ref{sec:appendix_site_energy_gap_stokes_shift}). With the energy-basis jump operators defined in Sec.~\ref{sec:model}, the state remains diagonal and the Lindbladian reduces exactly to a two-state Pauli master equation at $\omega_0=\Delta E/\hbar$. At $T=293$~K with $\Delta E = \sqrt{\Delta^2+4J^2}\approx90$~meV, detailed balance gives the equilibrium bright-state population $\rho_{++}^*=\gamma_{+-}/(\gamma_{-+}+\gamma_{+-})=1/(1+e^{-\Delta E/(k_BT)})\approx0.97$  (see Fig.~\ref{fig:energy_mixed_state_dynamics}), excluding pure dephasing as defined in Sec. \ref{sec:model}. The presence of an overwhelming $\rho_{++}^*\approx0.97$ fixed point could explain dimer brightness being about double that of the monomer within experimental error despite the presence of dark states. Furthermore, a true mixed state of bright and dark exciton dimers would result in multiple pathways for fluorescent emission, as the dark state would also be optically accessible once coherence is lost. Hence, even if the presence of antiparallel dipoles flips the energy ordering of the bright and dark states, fluorescent emission could still be thermally stabilized by strong excitonic coupling. A more detailed alternative microscopic description of dynamics, however, especially one resulting in longer-lived coherence, likely provides a somewhat altered, more nuanced picture.

\subsection{On Thermal Transfer Rates and the Effects of Localization}

The transfer rates of previously studied mVenus and mVenus-20nm-TD constructs \cite{Kim2019_VenusDimers} can be modeled as isotropic and using "classical" F\"orster resonance energy transfer (FRET) theory (see Appendix \ref{app:rates}). Here FRET rates of these two cases are compared to the thermal transfer rates predicted for the sub-ps and long-timescale dVenus exciton dimer models described above (see Tab. \ref{tab:rates} for rates, Sec. \ref{sec:results_antibunching} for sub-ps dynamical results, and Sec. \ref{sec:results_mixed} for long-timescale dynamical results).
\begin{table}[h!]
    \centering
    \begin{tabular}{|c|c|}\hline
         $k_{DA} (R_{DA}=2.5\text{ nm})$& 0.21 ps$^{-1}$   \\
         $k_{DA} (R_{DA}=20\text{ nm})$& $8.1\times10^{-8}$ ps$^{-1}$   \\
         $T_1^{-1}(0)$& 0.024 ps$^{-1}$  \\
         $T_1^{-1}(\infty)$& 2.20 ps$^{-1}$ \\\hline
    \end{tabular}
    \caption{\textbf{FRET and thermal transfer rates of comparable Venus systems.} Drastically faster long-timescale thermal transfer  rates $T_1^{-1}$ of the dVenus exciton dimer relative to isotropic FRET rates $k_{DA}$ of monomeric pair and 20nm-TD constructs can explain observed depolarization effects previously experimentally measured between the three constructs \cite{Kim2019_VenusDimers}. See Appendix \ref{app:rates} for details on FRET.}
    \label{tab:rates}
\end{table}

FRET rates were calculated using Venus-like values; $R_0=5.0$ nm, and $\tau_D=3.0$ ns \cite{Borst2005,FPbaseVenus} (see Appendices \ref{app:rates} and \ref{sec:appendix_parameters}). Drastically faster long-timescale thermal transfer of the dVenus exciton dimer relative to FRET rates (see Tab. \ref{tab:rates}) can explain observed depolarization effects previously experimentally measured between the three constructs \cite{Kim2019_VenusDimers}. Furthermore, increased distance between sites $r_{12}=R_{DA}$ also results in weaker coupling (see Eq. \ref{eq:PDA_J} and Appendix \ref{app:rates}), ultimately resulting in higher localization of excitations, causing relatively independent emission of sites observed in the mVenus-20nm-TD construct \cite{Kim2019_VenusDimers}.

The point--dipole approximation (PDA) can used to calculate the coupling energy $J$ between two dipoles at sites $\ket{1}$ and $\ket{2}$ as such \cite{Banerjee2025_DimerCoherence},
\begin{equation}
J_{12}^{\mathrm{PDA}} 
\propto \frac{1}{4\pi\varepsilon_0\,} 
  \left[
    \frac{\boldsymbol{\mu}_1 \cdot \boldsymbol{\mu}_2}{|r_{12}^3|}
    - 3 \frac{(\boldsymbol{\mu}_1 \cdot \mathbf{r}_{12})
              (\boldsymbol{\mu}_2 \cdot \mathbf{r}_{12})}{|r_{12}^5|}
  \right],
\label{eq:PDA_J}
\end{equation}
where $\varepsilon_0$ represents the vacuum permittivity, $\boldsymbol{\mu}_1$ and $\boldsymbol{\mu}_2$ represent the dipole moments of the respective site dipoles, \textbf{r}$_{12}$ represents the radial vector between them, and details of the dielectric environment \cite{Liebe1991,Gilmore2008QuantumDynamics,Li2013,Elton2017,BurgessFlorescu2024_arXiv} are excluded to emphasize the general form. Negative coupling energy $J$ observed in Venus dimers can hence occur due to typical J-dimer behavior \cite{Banerjee2025_DimerCoherence} or rather due to antiparallel, FP-like coupling geometry \cite{Wang2019} behavior, yielding a slightly different energetic landscape but identical values for simulated thermal transfer rates $T_1^{-1}$ (see Tab. \ref{tab:rates}). A drastically altered bath structure and change in system coupling strength could, however, considerably alter simulated thermal transfer rates, influencing quantitative and/or qualitative outcomes.

\section*{Conclusions}

Population dynamics of the Venus dimer were successfully modeled, yielding physically reasonable results with the caveat of strict violation of several assumptions underlying the approximations of a typical Lindblad framework. Nonetheless, complete trace-preservation and positivity are achieved on relevant timescales, and the resulting dephasing times and thermal transfer rates suggest meaningful insights into the mechanisms underlying the results of previous Venus experiments, originally published in 2019 \cite{Kim2019_VenusDimers}.

\section{Future Directions}
\label{sec:future_work}

Moving forward, bath structure should be altered with great care to add additional realism, such as modeling dynamical vibrational relaxation, and improved quantitative and/or qualitative accuracy and analytical interpretability of simulated dephasing times and thermal transfer rates, while satisfying any future chosen approximations. Furthermore, the additional FPs, whose analogous results were published this year \cite{Nguyen2025ExcitonicFPs}, particularly the improved measurement of the dVenus dimer's coupling strength, should be simulated, along with further analysis of FP structure--function relationship, to see if interpretations of the underlying coupling/absorption mechanism and the resulting dynamics can be significantly changed and/or deepened.

\section*{Acknowledgments}
The author acknowledges use of the QuTiP framework (v5.2.0) \cite{Johansson2012_QuTiP,Johansson2013_QuTiP2} for open quantum system simulations, NumPy (v2.3.2) \cite{Harris2020_NumPy} for array programming and scientific computing, SciPy (v1.16.0) \cite{Virtanen2020_SciPy} for scientific algorithms and numerical routines, and open-source PyMOL (v2.5, commit d24468af) \cite{PyMOL_OpenSource} for molecular graphics, and ChatGPT (4o and 5) (OpenAI) for assistance with text editing and with the generation and debugging of code. All information and code were critically reviewed and validated by the author.

I am grateful to J. McFadden for being my first contact at the University of Surrey and to Y. Kim for welcoming me as a PhD student and for granting the intellectual freedom to pursue the work presented here. I am also grateful to A. Rocco and M. Freed for introducing me to the world of theoretical physics research and to A. P. Kalra for encouraging me to pursue my own ideas.

I additionally thank several faculty members at Ursinus College for supporting my diverse interests and, in particular, R. Martin-Wells for keeping me on my toes throughout PHYS-207W (Modern Physics), where my foundation in quantum physics was originally built.

\section*{Funding}
This research received no external funding and was conducted using the author’s own time and computational resources.

\section*{Code and Data Availability}
All figure scripts, the conda environment, and supporting materials required to reproduce this work are archived on Zenodo, version v2.0.0-arxiv (DOI: 10.5281/zenodo.16892122) \cite{Abrahams2025_ZenodoFigures} and mirrored on GitHub (\url{https://github.com/ianthomasabrahams/venus-dimer-theory-figures}).

\section*{Conflict of Interest}
The author declares no conflict of interest.

\bibliographystyle{unsrt}
\bibliography{arxiv_submission_venus-dimer_v5}

\section*{Appendices: Parameters, Calculations, Definitions, and Derivations}
\appendix

\section{Simulation Physical Constants/System and Bath Parameters}
\label{sec:appendix_parameters}

Parameters for the system and bath and physical constants used in simulations are found below in Tab. \ref{tab:parameters}.

\begin{table}[h!]
  \centering
  \begin{tabular}{|c|c|c|}\hline
     Symbol&Labels& Value\\\hline
     Constants&CODATA 2022& \cite{RevModPhys.97.025002}\\\hline
     $\hbar$&Reduced Planck constant&0.6582119514 meV$\cdot$ps\\
     $k_B$&Boltzmann constant&0.08617333262 meV$\cdot$K$^{-1}$\\\hline
     Quantum&Lindblad&\cite{Breuer2002}\\\hline
    $\Delta$& Venus Effective Stokes shift& 59 meV \cite{Nagai2002Venus,FPbaseVenus}\\
    $|J|$& Coulombic coupling energy & 34 meV \cite{Kim2019_VenusDimers}\\
    $T$& Bath temperature&293 K \cite{Kim2019_VenusDimers}\\
    $\lambda$& Effective bath reorganization energy & 270 cm$^{-1}$ ($\sim 33.5$ meV) \cite{BurgessFlorescu2024_arXiv,Cinelli2001_PRL,Gilmore2008QuantumDynamics}\\
    $\tau_c$&Bath memory time& 100 fs (0.1 ps) \cite{Hutchison2023UltrafastFP}\\
    $\gamma_c$& Bath memory cutoff rate& 10 ps$^{-1}$ (1/$\tau_c$)\\\hline
    Classical&FRET&\cite{Clegg1996_FRET}\\\hline
    $\tau_D$& Venus fluorescence lifetime & 3.0 ns \cite{FPbaseVenus}\\
    $R_0$& GFP F\"orster radius & 5.0 nm \cite{Borst2005}\\\hline
  \end{tabular}
  \caption{Model parameter symbols, labels, and numerical values used in simulations.}
  \label{tab:parameters}
\end{table}

\section{Energy Gap Calculations, Unit Conversions, and Site-Basis Rotation}
\subsection{Conversion of Stokes Shift from Wavelength to Energy}
\label{sec:appendix_site_energy_gap_stokes_shift}

In fluorescence spectroscopy, the Stokes shift is typically reported as the difference in wavelength between the absorption maximum $\lambda_{\text{abs}}$ and the emission maximum $\lambda_{\text{em}}$. To express this shift in energy units, the Stokes shift in eV is computed,
\begin{equation}
  \Delta 
  = hc\left(\frac{1}{\lambda_{\text{abs}}} - \frac{1}{\lambda_{\text{em}}}\right),
\end{equation}
where $h$ is Planck's constant and $c$ is the speed of light. 

For numerical evaluation, $hc = 1239.84\ \text{eV}\cdot\text{nm}$, with $\lambda_{\text{abs}}$ and $\lambda_{\text{em}}$ in nanometers. For example, in the case of Venus, $\lambda_{\text{abs}} = 515$~nm, and $\lambda_{\text{em}} = 528$~nm \cite{Nagai2002Venus}, hence 
\begin{align}
  \Delta 
  &= 1239.84\ \text{eV}\cdot\text{nm}\,\left(\frac{1}{515\ \text{nm}} - \frac{1}{528\ \text{nm}}\right) \\
  &\approx 0.0593\ \text{eV} \\
  &\boxed{\Delta = 59\ \text{meV}}
\end{align}

\noindent

\subsection{Conversion of Davydov Splitting from Wavelength to Energy}
\label{sec:appendix_splitting}

Previously, Kim \emph{et al.} reported a Davydov splitting of $\Delta\lambda = 14.6 \pm 0.3\ \text{nm}$ centered at $\lambda_0 \approx 516\ \text{nm}$, as observed in the circular dichroism spectra of dVenus dimers \cite{Kim2019_VenusDimers}. To convert this spectral splitting into an energy gap, the relation between photon energy and wavelength is used:

\begin{equation}
 2|J|
 = hc\left(\frac{1}{\lambda_{\text{max}}} - \frac{1}{\lambda_{\text{min}}}\right),
\end{equation}
where $h$ is Planck's constant and $c$ is the speed of light. 

For numerical evaluation, $hc = 1239.84\ \text{eV}\cdot\text{nm}$ \cite{RevModPhys.97.025002}, with $\lambda_{\text{max}}$ and $\lambda_{\text{min}}$ in nanometers. For example, in the case of Venus, $\lambda_{\text{max}} = 507.4$~nm, and $\lambda_{\text{min}} = 522$~nm \cite{Nagai2002Venus}, hence 
\begin{align}
 2|J| 
 &= 1239.84\ \text{eV}\cdot\text{nm}\,\left(\frac{1}{507.4\ \text{nm}} - \frac{1}{522\ \text{nm}}\right) \\
 &\approx 0.0683\ \text{eV} \\
 &2|J| \approx68\ \text{meV }\\
 &\boxed{|J|=34\text{ meV}}
\end{align}

\subsection{Rotation to the Site Basis}
\label{sec:appendix_rotation_site_basis_Ham}

Transforming from the excitonic basis to the site basis introduces off-diagonal coherence
that couples site populations through the delocalized eigenstates. For a site-energy difference $\Delta$ (Stokes shift) and coupling $J$, the Hamiltonian in the site basis is hence
\begin{equation}
\hat{H} = \begin{pmatrix}
\Delta/2 & J \\
J & -\Delta/2
\end{pmatrix}\qquad\to\text{diagonalization} \to \qquad\hat{H}=\begin{pmatrix}
  E_+&0\\
  0&E_-
\end{pmatrix}.
\end{equation}
Diagonalization yields eigenenergies $E_\pm = \pm \tfrac{1}{2}\Delta E$, where $\Delta E = \sqrt{\Delta^2+4J^2}$. In this representation, population relaxation and dephasing are controlled by $\gamma_\phi$ together with the thermal transfer rates $\gamma_{+-}$ and $\gamma_{-+}$, which redistribute exciton populations according to detailed balance. This site-basis perspective is essential for analyzing relaxation pathways following structural or solvent-induced energy shifts (Appendix \ref{sec:appendix_rates}).

\section{Time-Step and Window Selection}
\label{sec:appendix_time_step_window}
The simulation time step $\Delta t$ and total duration $T$ are chosen from physical timescales of the dimer.

\paragraph{Fastest oscillation.}
For excitonic coupling $J$ (in meV), the exciton splitting is
\[
\Delta E = 2|J|.
\]
With $\hbar = 0.658\,212~\text{meV}\cdot\text{ps}$, the angular frequency and period of the fastest coherent oscillation are
\[
\omega_{\max}=\frac{\Delta E}{\hbar}, \qquad
T_{\text{osc}}=\frac{2\pi}{\omega_{\max}}.
\]
For $J=-34$~meV, $\Delta E=68$~meV so
\[
\omega_{\max}=\frac{68}{0.658212}\;\text{ps}^{-1}\approx 1.033\times 10^{2}\ \text{ps}^{-1},
\qquad
T_{\text{osc}}\approx 6.082\times 10^{-2}\ \text{ps}=60.8~\text{fs}.
\]

\paragraph{Time step (Nyquist + oversampling).}
To resolve oscillations at frequency $\omega_{\max}$, the sampling theorem requires
$\Delta t < \pi/\omega_{\max}$ \cite{Press2007_NumericalRecipes}. 

The dynamics are oversampled by choosing
\[
\Delta t=\frac{T_{\text{osc}}}{N_{\text{spp}}},
\]
with $N_{\text{spp}}\in[100,200]$ samples per period (used here: $N_{\text{spp}}=200$), giving
\[
\Delta t \approx \frac{60.8~\text{fs}}{200}\approx 0.304~\text{fs}.
\]
This also satisfies stability for the Lindblad solver when dissipative rates are
$\gamma \ll 1/\Delta t$.

\paragraph{Total duration (frequency resolution / decay capture).}
To resolve a minimum frequency (linewidth) scale $\delta f$ or a slow decay rate,
$T\gtrsim 1/\delta f$ is required. Equivalently, to observe dephasing with half-life
$t_{1/2}$ over several e-folds, a decay rate is taken such that
\[
T \approx (6\text{--}10)\,t_{1/2}.
\]
In the figures, independent windows per panel were used, sized to the temperature-dependent
$t_{1/2}$ (colder $\Rightarrow$ longer $T$), with the common time step $\Delta t$ above.

\noindent\textbf{Remark.} For effective heterodimers with a Stokes shift $\Delta\neq 0$, the fastest oscillation is set by
$\Delta E=\sqrt{\Delta^2+4J^2}$ (not $2|J|$), and apply the same $\Delta t=T_{\mathrm{osc}}/N_{\mathrm{spp}}$ rule.

\section{Dynamics in a Drude-Lorentz Bath: Effective Reorganization Energy, Dephasing, and Thermal Transfer/Relaxation}
\label{sec:appendix_rates}
\textbf{*Note:} see Appendix \ref{sec:appdx_deriv_thermal_transfer_rates} for detailed derivations.

\subsection{Effective Reorganization Energy from Experimental Coherence Times}
\label{sec:appendix_lambda_fast}

Cinelli \emph{et al.}\ reported room-temperature coherent vibrational dynamics in GFP with a \emph{dephasing time of about 1 ps} \cite{Cinelli2001_PRL}. Following Burgess and Florescu’s double-Debye dielectric model for the GFP chromophore pocket (cavity geometry plus two solvent relaxation channels) \cite{BurgessFlorescu2024_arXiv}, the pure-dephasing rate obeys the general Bloch--Redfield relation
\begin{equation}
\gamma_\phi \;=\; \pi\,T\;\lim_{\omega\to 0}\frac{J(\omega)}{\omega},
\end{equation}
with \(T \equiv k_B T/hc \approx 208~\mathrm{cm}^{-1}\) at 293~K.

Evaluating this expression with $T_2=1/\gamma_\phi \approx 1~\mathrm{ps}$ as a heuristic dephasing time and using the double-Debye spectral density yields an effective reorganization
\begin{equation}
\boxed{\;\lambda = 2.7\times10^{2}~\mathrm{cm}^{-1}\approx 33.5 \text{meV}\;}
\end{equation}
used for all simulations, consistent with the ps-scale dynamics reported by Cinelli \emph{et al.} as well as previous order-of-magnitude estimates for FP reorganization energies \cite{BurgessFlorescu2024_arXiv,Gilmore2008QuantumDynamics}.

\subsection{Pure Dephasing in the Energy Basis}
\label{sec:appendix_dephasing_rate}

Consider a system with Hamiltonian $\hat{H}_S$ that is linearly and diagonally coupled to a harmonic bath,
\begin{equation}
\hat{H} = \hat{H}_S + \sum_\xi \hbar\omega_\xi\, \hat{b}^\dagger_\xi \hat{b}_\xi + \hat{A} \otimes \sum_\xi c_\xi (\hat{b}_\xi + \hat{b}^\dagger_\xi),
\end{equation}
with $[H_S,A]=0$. In the cumulant (Kubo) treatment for a Gaussian bath, the coherence between energy eigenstates $\ket{+}$ and $\ket{-}$ decays as \cite{Kubo1966,Breuer2002}
\begin{equation}
\rho_{+-}(t)=\rho_{+-}(0)\,e^{-i\omega_0 t}\,e^{-g(t)},\qquad 
g(t)=\int_{0}^{t}\!d\tau\int_{0}^{\tau}\!d\tau'\, C_\omega(\tau'),
\end{equation}
where $C_\omega(t)=\langle \delta\omega(t)\,\delta\omega(0)\rangle$ with $\delta\omega(t)=\delta E(t)/\hbar$.

For a Drude-Lorentz spectral density
\begin{equation}
J(\omega)=\frac{2\lambda\,\omega\,\gamma_c}{\omega^2+\gamma_c^2},
\end{equation}
where $\gamma_c = \tau_c^{-1}$, and $\tau_c$ is the bath correlation time, the energy-gap correlation function is
\begin{equation}
C_E(t)=\lambda\gamma_c\!\left[\coth\!\left(\tfrac{\beta\hbar\gamma_c}{2}\right)-i\right]e^{-\gamma_c t}+\text{Matsubara terms}.
\end{equation}
In the classical/high-temperature limit, Matsubara terms can be neglected and
\begin{equation}
C_\omega(t)=\frac{1}{\hbar^2}\,\mathrm{Re}\,C_E(t)\simeq 
\frac{2\lambda k_B T}{\hbar^2}\,e^{-\gamma_c t}.
\end{equation}
The lineshape function evaluates to
\begin{equation}
g(t)=\frac{2\lambda k_B T}{\hbar^2}\left[\frac{t}{\gamma_c}-\frac{1-e^{-\gamma_c t}}{\gamma_c^2}\right],
\end{equation}
with limits
\[
g(t)\simeq \frac{\lambda k_B T}{\hbar^2}t^2\quad (t\ll \tau_c),\qquad
g(t)\simeq \frac{2\lambda k_B T}{\hbar^2\gamma_c}\,t\quad (t\gg \tau_c).
\]
Thus, in the motional-narrowing limit the pure-dephasing rate used is
\begin{equation}
\boxed{\;\gamma_\phi = \frac{2\lambda k_B T}{\hbar^2\gamma_c}\;}
\label{eq:gamma_phi}
\end{equation}

\medskip
\noindent\textbf{Remark on correlation time.} 
Where the single-Debye Drude-Lorentz form [Eq.~\eqref{eq:gamma_phi}] is invoked, conservative correlation time $\tau_c = 0.1~\mathrm{ps}$ is adopted, aligned with reasonable timescales for FP systems \cite{Hutchison2023UltrafastFP}.

\subsection{Thermal Population Transfer and Relaxation (Born--Markov--secular Limit)}
\label{sec:appendix_thermal_transfer}

Within the Born--Markov--secular approximation \cite{Breuer2002}, thermal transitions between excitonic states
are described by downhill ($\gamma_{-+}$) and uphill ($\gamma_{+-}$) rates. These follow from
Fermi’s golden rule applied to the bath spectral density, with detailed balance ensuring
\begin{equation}
\frac{\gamma_{+-}}{\gamma_{-+}} = e^{-\beta \Delta E},
\end{equation}
where $\Delta E = \sqrt{\Delta^2+4J^2}=\hbar\omega_0$ is the exciton splitting, with site energy difference $\Delta$ and coupling energy $J$ previously defined in Sec. \ref{sec:model}, and inverse temperature $\beta = 1/k_B T$. The downhill rate $\gamma_{-+}$ is set by the overlap of the spectral density with the excitonic transition frequency, while the uphill rate $\gamma_{+-}$ is exponentially suppressed at low temperature. Together with the pure-dephasing rate $\gamma_\phi$, these thermal rates define the full Lindblad dynamics of the dimer, governing both relaxation into the bright state and the loss of coherence between excitons.

Next the derivation of the Drude--Lorentz pure dephasing, as defined above, is shown.

\section{Detailed Derivation of Thermal Transfer and Pure Dephasing Rates in a Drude--Lorentz Bath}
\label{sec:appendix_derivations_therm+deph}

In this section we will walk through a step-by-step derivation of the Born--Markov--secular 
thermal transfer rates and Drude--Lorentz pure dephasing rates, and explicitly 
demonstrate their internal consistency. The goal is to make transparent the 
approximations and physical intuition, so that readers new to open quantum systems 
can follow the logic.

\subsection*{System--Bath Hamiltonian}

We begin with a two-level system (the excitonic dimer) coupled linearly to a 
harmonic bath:
\begin{equation}
\hat{H} = \hat{H}_S + \hat{H}_B + \hat{H}_{SB},
\end{equation}
where
\begin{align}
\hat{H}_S &= \sum_{\alpha=\pm} E_\alpha \, |\alpha\rangle\langle \alpha|, \\
\hat{H}_B &= \sum_\xi \hbar \omega_\xi \, \hat{b}_\xi^\dagger \hat{b}_\xi, \\
\hat{H}_{SB} &= \hat{A} \otimes \sum_\xi c_\xi (\hat{b}_\xi + \hat{b}_\xi^\dagger).
\end{align}
Here $|\alpha\rangle$ denotes excitonic eigenstates with energies $E_\pm$, 
$\hat{b}_\xi^\dagger, \hat{b}_\xi$ are bosonic creation/annihilation operators, 
and $\hat{A}$ is the system operator coupling to bath fluctuations.

\subsection*{Spectral Density and Drude--Lorentz Model}

The influence of the bath is fully encoded in the spectral density,
\begin{equation}
J(\omega) = \sum_\xi c_\xi^2 \, \delta(\omega - \omega_\xi).
\end{equation}
For condensed-phase environments such as proteins, a widely used form is the 
Drude--Lorentz spectral density:
\begin{equation}
J(\omega) = \frac{2 \lambda \, \omega \, \gamma_c}{\omega^2 + \gamma_c^2},
\end{equation}
where $\lambda$ is the reorganization energy (bath coupling strength) and 
$\gamma_c$ is the cutoff frequency (inverse bath memory time).

\subsection*{Thermal Transfer Rates (Born--Markov--secular)}
\label{sec:appdx_deriv_thermal_transfer_rates}
In the Born--Markov--secular limit, energy relaxation between excitonic states 
is obtained from Fermi’s golden rule:
\begin{equation}
\gamma_{+-} = 2\pi J(\omega_0)\,[n_{BE}(\omega_0,T)+1]\text{ (downhill/favored rate)}, \quad 
\gamma_{-+} = 2\pi J(\omega_0)\,n_{BE}(\omega_0,T)\text{ (uphill/suppressed rate)},
\end{equation}
with transition frequency $\omega_0 = (E_+ - E_-)/\hbar$ and Bose--Einstein 
occupancy factor
\begin{equation}
n_{BE}(\omega,T) = \frac{1}{e^{\hbar\omega/k_BT} - 1}.
\end{equation}
Inserting the Drude--Lorentz form yields the ``classical'' spectral prefactor
\begin{equation}
S_{cl}(\omega_0) = \frac{2\lambda k_B T}{\hbar^2} 
\frac{\gamma_c}{\gamma_c^2+\omega_0^2},
\end{equation}
so that
\begin{equation}
\gamma_{+-} = S_{cl}(\omega_0)\,[n_{BE}(\omega_0,T)+1], \qquad
\gamma_{-+} = S_{cl}(\omega_0)\,n_{BE}(\omega_0,T).
\end{equation}
These rates obey detailed balance:
\begin{equation}
\frac{\gamma_{-+}}{\gamma_{+-}} = e^{-\hbar \omega_0 / k_B T}.
\end{equation}

\subsection*{Pure Dephasing (Cumulant Expansion)}

Next we examine coherence decay between $|+\rangle$ and $|-\rangle$. 
For diagonal system--bath coupling ($[\hat{H}_S,\hat{A}]=0$), the off-diagonal 
density matrix element evolves as
\begin{equation}
\hat{\rho}_{+-}(t) = \hat{\rho}_{+-}(0)\, e^{-i\omega_0 t} e^{-g(t)},
\end{equation}
with lineshape function
\begin{equation}
g(t) = \int_0^t d\tau \int_0^\tau d\tau' \, C_\omega(\tau'),
\end{equation}
where $C_\omega(t)$ is the energy-gap correlation function. 
For the Drude--Lorentz bath, in the classical/high-temperature limit:
\begin{equation}
C_\omega(t) \simeq \frac{2\lambda k_B T}{\hbar^2} e^{-\gamma_c t}.
\end{equation}
Evaluating the integrals gives
\begin{equation}
g(t) = \frac{2\lambda k_B T}{\hbar^2}
\left( \frac{t}{\gamma_c} - \frac{1-e^{-\gamma_c t}}{\gamma_c^2} \right).
\end{equation}

\subsection*{Short- and Long-Time Limits}

\paragraph{Short-time expansion ($t \ll \tau_c$):}
\begin{equation}
g(t) \approx \frac{\lambda k_B T}{\hbar^2}\, t^2, 
\end{equation}
implying Gaussian coherence decay, as expected from a cumulant treatment. 
The instantaneous dephasing rate behaves as
\begin{equation}
\gamma_\phi(t) \approx \frac{2\lambda k_B T}{\hbar^2}\, t.
\end{equation}

\paragraph{Long-time Markovian limit ($t \gg \tau_c$):}
\begin{equation}
g(t) \approx \frac{2\lambda k_B T}{\hbar^2 \gamma_c}\, t,
\end{equation}
so that exponential decay is recovered with a constant pure-dephasing rate:
\begin{equation}
\gamma_\phi = \frac{2\lambda k_B T}{\hbar^2 \gamma_c}.
\end{equation}

\subsection*{Internal Consistency with Thermal Transfer Rates}

Note that $\gamma_\phi$ arises from the \emph{zero-frequency limit} of the 
same spectral density that controls thermal transfer. Specifically,
\begin{equation}
\lim_{\omega_0 \to 0} S_{cl}(\omega_0) = 
\frac{2\lambda k_B T}{\hbar^2 \gamma_c} = \gamma_\phi.
\end{equation}
Thus, pure dephasing is simply the $\omega_0 \to 0$ (elastic scattering) 
continuation of the inelastic thermal transfer rates. Both are governed by the 
same bath parameters $(\lambda, \gamma_c, T)$, confirming their consistency.

\subsection*{Summary}

\begin{itemize}
  \item Thermal population transfer rates $\gamma_{+-}, \gamma_{-+}$ arise from 
  bath-induced transitions at frequency $\omega_0$, obeying detailed balance.
  \item Pure dephasing $\gamma_\phi$ arises from low-frequency bath 
  fluctuations, with Gaussian decay at short times and exponential decay in the 
  Markovian limit.
  \item Both are derived from the same Drude--Lorentz spectral density, and 
  $\gamma_\phi$ is recovered as the $\omega_0 \to 0$ limit of the thermal 
  transfer spectral prefactor $S_{cl}(\omega_0)$.
\end{itemize}
This unification shows that relaxation and dephasing are two aspects of the same 
system--bath interaction, differing only by the relevant frequency window of the 
bath.

\subsection*{Beyond the High-Temperature, Markovian Approximation}

The preceding derivation illustrated the essence of how pure dephasing emerges 
from a Drude--Lorentz bath, but it relied on two key simplifications: 
(i) the high-temperature limit $k_BT \gg \hbar \gamma_c$, and 
(ii) the long-time Markovian limit $t \gg \tau_c$. 
Here we extend the treatment to show the more general form, retaining both 
finite-temperature statistics and non-Markovian time dependence.

\paragraph{Full bath correlation function.}
Starting again from the Drude--Lorentz spectral density,
\begin{equation}
J(\omega) = \frac{2\lambda \omega \gamma_c}{\omega^2 + \gamma_c^2},
\end{equation}
the corresponding energy-gap correlation function is
\begin{equation}
C_E(t) = \lambda \gamma_c 
\left[\coth\!\left(\frac{\beta \hbar \gamma_c}{2}\right) - i\right] e^{-\gamma_c t} 
+ \text{Matsubara terms},
\end{equation}
where $\beta = 1/(k_B T)$. 
The real part governs fluctuations (dephasing), while the imaginary part leads 
to renormalization of system energies. 
The Matsubara terms provide quantum corrections to the high-temperature form 
and decay rapidly at room temperature.

\paragraph{Classical/high-temperature limit.}
Neglecting Matsubara terms and taking 
$\coth(\beta \hbar \gamma_c/2) \approx 2k_B T/(\hbar \gamma_c)$, one obtains
\begin{equation}
C_\omega(t) \simeq \frac{2\lambda k_B T}{\hbar^2} e^{-\gamma_c t}.
\end{equation}
This reproduces the simplified correlation function used earlier.

\paragraph{Non-Markovian lineshape function.}
Evaluating the cumulant integral with this exponential correlation function 
yields
\begin{equation}
g(t) = \frac{2\lambda k_B T}{\hbar^2}
\left(\frac{t}{\gamma_c} - \frac{1-e^{-\gamma_c t}}{\gamma_c^2}\right).
\end{equation}
Thus the instantaneous pure-dephasing rate is
\begin{equation}
\gamma_\phi(t) = \frac{2\lambda k_B T}{\hbar^2 \gamma_c}
\left(1-e^{-\gamma_c t}\right).
\end{equation}
This form captures both the non-Markovian short-time regime 
(Gaussian decay with $\gamma_\phi(t)\propto t$) and the long-time Markovian 
limit $\gamma_\phi = \tfrac{2\lambda k_B T}{\hbar^2 \gamma_c}$.

\paragraph{Site correlation correction.}
For two strongly coupled chromophores sharing a structured environment, 
fluctuations may be highly correlated \cite{Rolczynski2018CorrelatedProteinEnvironments}. 
To account for this, the effective pure-dephasing rate is scaled by
\begin{equation}
\gamma_\phi(t) = (1-c_s)\,
\frac{2\lambda k_B T}{\hbar^2 \gamma_c}\,
\left(1-e^{-\gamma_c t}\right),
\end{equation}
with $c_s \in [0,1]$ denoting the degree of site correlation. 
The limit $c_s \to 1$ represents nearly identical site environments, strongly 
suppressing dephasing.

\paragraph{Summary of corrections.}
Relative to the simplified treatment, the refinements here introduce:
\begin{itemize}
  \item \emph{Finite-temperature statistics:} explicit $\coth(\beta \hbar \gamma_c/2)$ 
  factor, reducing to $2k_B T/(\hbar \gamma_c)$ at high $T$.
  \item \emph{Non-Markovianity:} explicit exponential memory factor 
  $1-e^{-\gamma_c t}$ describing the crossover from Gaussian short-time decay 
  to exponential long-time decay.
  \item \emph{Correlated environments:} site correlation constant $c_s$ that 
  interpolates between uncorrelated ($c_s=0$) and fully correlated ($c_s=1$) 
  dephasing dynamics.
\end{itemize}

Together, these corrections show how the simplified expressions given previously 
are recovered as limiting cases of the more general non-Markovian treatment.

\subsection*{Correlation Effects and the Origin of the $(1-c_s)$ Factor}

So far we have treated dephasing as if the two sites of the dimer couple to 
independent baths. In reality, closely interacting chromophores embedded in the 
same protein scaffold may experience highly correlated fluctuations. This 
modifies the effective dephasing rate. We now derive the $(1-c_s)$ factor from 
the general correlated-bath picture.

\paragraph{General system--bath interaction.}
Let each site $i=1,2$ couple diagonally to the bath via operators 
$\hat{A}_i = |i\rangle\langle i|$, so that
\begin{equation}
\hat{H}_{SB} = \sum_{i=1}^2 \hat{A}_i \otimes \hat{B}_i,
\end{equation}
with $\hat{B}_i$ bath operators. 
The corresponding bath correlation functions are
\begin{equation}
C_{ij}(t) = \langle \hat{B}_i(t)\,\hat{B}_j(0)\rangle,
\end{equation}
with $i,j \in \{1,2\}$. 
These include both auto-correlations ($i=j$) and cross-correlations ($i\neq j$).

\paragraph{Decoherence of inter-site coherence.}
For a superposition of site states, e.g.\ 
$(|1\rangle - |2\rangle)/\sqrt{2}$, the relevant coherence is 
$\rho_{12}(t) = \langle 1|\hat{\rho}(t)|2\rangle$. 
Its decay rate depends on the fluctuation of the \emph{difference} operator 
$\hat{A}_1 - \hat{A}_2$. 
The corresponding noise correlation is
\begin{equation}
\tiny\langle \big(\hat{B}_1(t)-\hat{B}_2(t)\big)
    \big(\hat{B}_1(0)-\hat{B}_2(0)\big)\rangle
= C_{11}(t) + C_{22}(t) - C_{12}(t) - C_{21}(t).
\end{equation}

\paragraph{Defining the correlation coefficient.}
If both sites couple to baths of equal strength, then 
$C_{11}(t)=C_{22}(t)=C(t)$. 
We can then define the correlation coefficient $c_s$ as the \emph{ratio} of 
cross- to auto-correlations:
\begin{equation}
c_s = \frac{C_{12}(t)}{C_{11}(t)} = \frac{C_{21}(t)}{C(t)}.
\end{equation}
By construction, $c_s \in [0,1]$: $c_s=0$ corresponds to independent baths, 
while $c_s=1$ corresponds to perfectly correlated baths.

\paragraph{Effective correlation driving dephasing.}
Substituting this into the effective noise correlation gives
\begin{equation}
C_{\text{eff}}(t) = 2\big(1-c_s\big)C(t).
\end{equation}
Thus the fluctuations that actually drive inter-site dephasing are reduced by 
the factor $(1-c_s)$.

\paragraph{Resulting pure-dephasing rate.}
Therefore, the general non-Markovian pure-dephasing rate becomes
\begin{equation}
\gamma_\phi(t) = (1-c_s)\,
\frac{2\lambda k_B T}{\hbar^2 \gamma_c}\,
\left(1-e^{-\gamma_c t}\right).
\end{equation}
The factor $(1-c_s)$ arises naturally from cross-correlations between 
site-specific baths: if $c_s=1$, fluctuations are perfectly correlated and the 
relative phase is preserved (no dephasing), while if $c_s=0$, the sites fluctuate 
independently and the full dephasing rate is recovered.

\paragraph{Physical analogy.}
\emph{An intuitive picture is provided by two pendulums hanging from the same 
beam. If the beam is shaken, both pendulums move together and their relative 
phase is unaffected ($c_s=1$). If each pendulum is buffeted by independent gusts 
of wind, their relative phases wander independently ($c_s=0$). Real systems fall 
between these extremes, with partial correlation $(0<c_s<1)$, leading to 
suppressed but nonzero dephasing.}

\section{On Classical vs. Quantum Transfer Rates}
\label{app:rates}

In the present regime of strong coupling, classical approximations such as F\"orster resonance energy transfer (FRET) theory break down. The monomeric forms of FPs, such as mVenus, may be more accurately described using FRET theory due to their isotropic rotation and hence weaker/randomized coupling relative to site energy gap as compared to geometrically stable dimers. These differences in rates, coupling strengths, and relative dipole orientations provide a direct link to depolarization, such as that observed in the form of concentration-/dimerization-dependent limiting anisotropy by Kim \emph{et al.} (2019) and Jim\'enez \emph{et al.} (2023) \cite{Kim2019_VenusDimers,Jimenez2023Ultrafast}. To compare classical and quantum limits \cite{Nelson2018QuantumDecoherenceFRET,Clegg1996_FRET,Rebentrost2009}, FRET transfer rates (Eq.~\ref{eq:k_DA}) are evaluated alongside the thermally mediated excitonic relaxation rates introduced in Sec.~\ref{sec:model}, thereby redefining the localized site states and transfer rates of the system to introduce the contextual language and key definitions of FRET theory.

First, the donor--acceptor basis for a homodimer or identical monomer pair is defined as,
\begin{equation}
  \ket{D}=\ket{2},\qquad \ket{A}=\ket{1}.
\end{equation}
Next, the FRET rate is defined as
\begin{equation}
  k_{DA}=\frac{1}{\tau_D}\bigg(\frac{R_0}{R_{DA}}\bigg)^6,
  \label{eq:k_DA}
\end{equation}
where $\tau_D$ is the lifetime of the donor in the absence of an acceptor, $R_{DA}$ is defined here as the average distance between donor and acceptor, and the corresponding F\"orster radius is defined as
\begin{equation}
  R_0=\Bigg|\sqrt[6]{\frac{9000\,\text{ln}(10)\,\kappa^2\Phi_D}{128\pi^5 N_A n^5}}\Bigg|,
\end{equation}
where $\kappa^2$ is the relative dipole orientation factor, $\Phi_D$ is the donor quantum yield, $N_A$ is the number of acceptor chromophores, and $n^5$ is the protein refractive index \cite{Clegg1996_FRET}. Hence a classical benchmark is given against which to interpret the present excitonic Lindblad model of FP coupling dynamics, linked directly to experimental observables.

\section{Extended Discussion}
\label{sec:extended}

The results above establish a mechanistic framework for reconciling antibunched photon statistics, 
strong excitonic coupling, and spectral shifts in Venus dimers. Yet these findings also 
invite a broader perspective. Excitonic phenomena in FPs not only 
clarify the mechanistic basis of engineered Venus constructs, but also reveal how various classes of CP may have more generally evolved to exploit quantum mechanical effects in service of biological function. In this sense, FP dimers can be viewed as both a tractable model of adaptive design in nature and a platform for emerging quantum technologies. 

\textbf{*Note:} the following subsections represent speculative explorations based on unknown frontiers of knowledge. The concepts explored and developed below are not founded strongly based on the results in the main text of this documents and should be regarded as a mental exercise in creative extrapolation. Furthermore, the J--H aggregate concept explored below, in particular, depends highly on specifics of relative dipole orientation and stands more as an information-theoretic concept than a true biological hypothesis.

\subsection{Evolutionary Implications for Chromoprotein Structure and Function}
\label{sec:evolution}

\begin{figure}[h!tbp]
  \centering
  \includegraphics[width=1\linewidth]{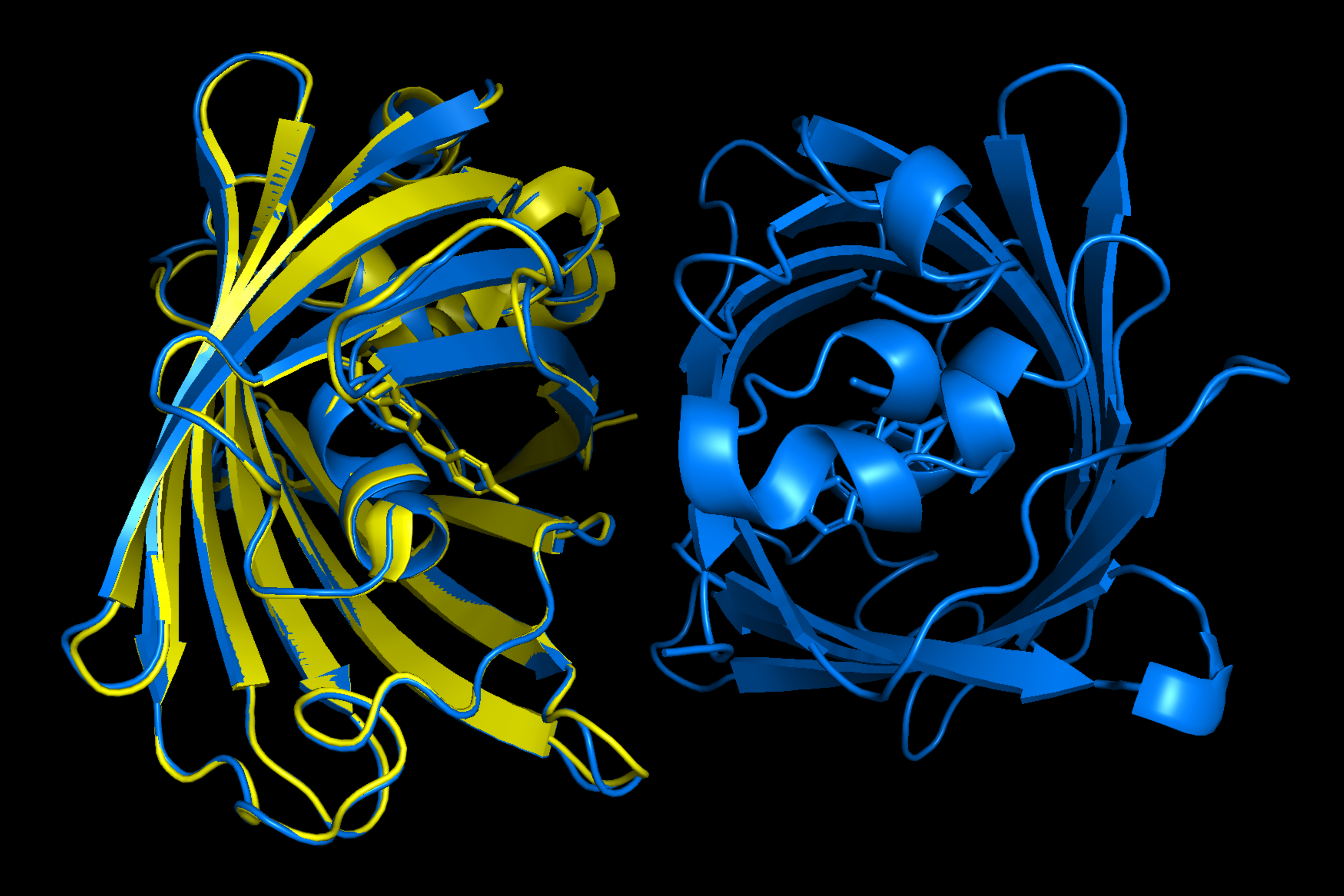}
  \caption{\textbf{Structural homology of Venus YFP and wild-type GFP.} 
  Global alignment of cartoon crystal structures of the Venus YFP subunit (yellow; PDB ID: 1MYW) and wild-type GFP dimer (marine blue; PDB ID: 1GFL) (RMSD: 0.323 \AA, computed in PyMOL global alignment) highlights conserved chromophore orientation relative to the $\beta$-barrel scaffold and overall structural homology.  (Chromophores are shown as sticks.) The roughly $C_2$-symmetric quaternary structure predicted by previous simulations of wild-type GFP dimerization \cite{Wang2019} is approximated by the corresponding crystal unit cell structure shown here.}
  \label{fig:1MYW_1GFL_aligned}
\end{figure}

Given the high structural conservation between wild-type GFP and Venus (Figure~\ref{fig:1MYW_1GFL_aligned}), it is reasonable to propose that thermal bright-state selection may represent a primary adaptive outcome of wild-type GFP's quaternary structure in \emph{A.~victoria}. This hypothesis, however, 
comes with important caveats to consider.

Molecular graphics and analysis were performed using open-source PyMOL (version 2.5, commit d24468af) \cite{PyMOL_OpenSource}, available at \href{https://github.com/schrodinger/pymol-open-source}{https://github.com/schrodinger/pymol-open-source}.

The PyMOL alignment (Figure~\ref{fig:1MYW_1GFL_aligned}) highlights strong structural 
similarity between Venus (1MYW) and wild-type GFP (1GFL), extending to the orientation of the chromophore axis within the $\beta$-barrel. This feature is particularly relevant, since excitonic coupling and fluorescence brightness in dimers are highly geometry dependent. The GFP crystal structure reflects the approximate $C_2$ symmetry expected of weak FP dimers \cite{Wang2019}, which may underlie the J-like negative coupling energy $J$ observed in Venus dimers. Such conservation reinforces the use of Venus/GFP as comparable model systems for reasoning about the adaptive roles of naturally occurring \emph{Aequorea} FPs.

\textbf{The adaptive hypothesis.} Recent cloning and spectroscopy across \textit{Aequorea} spp.\ uncovered a wide diversity of structural homologs spanning absorption from the blue--green into the far-red, including unusually bright-green FPs and entirely non-fluorescent CPs \cite{Lambert2020_Aequorea,Shaner2020_ImagingData}. These proteins were sourced from two species: the well-studied \emph{A.~victoria} and the less explored but closely related \emph{A.~cf.~australis}. From the latter, Lambert \emph{et al.} identified a close avGFP-like ortholog, AausGFP---here nicknamed the \emph{jade fluorescent protein} (JFP) for its slightly blue-shifted green emission, distinguishing it from canonical GFP from \emph{A.~victoria} \cite{Lambert2020_Aequorea}. (Here, JFP is to AausGFP as GFP is to avGFP.) 

Both GFP and JFP share conserved residues, including the hydrophobic dimerization interface, and exhibit similar spectra with the canonical neutral/anionic dual peak ($\sim$400/480~nm), suggesting adaptive traits conserved through speciation. Notably, the relative intensity of these peaks differs between the species, hinting at differences in the thermal equilibrium populations of neutral vs.\ anionic chromophore states. Such differences may reflect adaptation to the distinct thermal ranges of their native habitats, where the relative coldness compared to the ``room-temperature'' conditions used in laboratory studies \cite{Kim2019_VenusDimers,Purcell2018Successes,Lambert2020_Aequorea} likely plays a significant role in shaping realistic wild-type dynamics. 

It is therefore natural to propose that, among the many FP variants in \emph{Aequorea} spp., GFP and JFP stand out as the strongest candidates for functionally adaptive structures/sequences. By extension, as a GFP mutant, Venus provides a likely valid proxy, suggesting that thermal bright-state selection (Figure~\ref{fig:energy_mixed_state_dynamics}) may represent an adaptive trait emerging from strong excitonic coupling---even in the absence of long-lived quantum coherence. The full evolutionary picture is undoubtedly more complex, as will soon be pointed out, yet Venus remains a useful analogy for exploring the structural and functional bases of GFP and JFP adaptation.

\textbf{Signaling functionality and BRET constraints.} A classic retrodictive hypothesis is that \textit{A.~victoria} GFP’s spectra are constrained by the need to accept bioluminescent energy from its native BRET donor, aequorin (peak emission $\sim$475~nm), and to emit green underwater as a behavioral signal, such as to attract prey or deter predators \cite{Gorokhovatsky2004_BRET,Cubitt1995GFP}. Fluorescence may, however, also serve multiple behavioral functions, potentially including sexual selection, where visual signals are common across the Animal Kingdom \cite{Darwin1871Descent}. 

Aequorin fusion experiments indicate that GFP’s anionic chromophore is favored during interactions with calcium-bound aequorin, manifested as a redshift in GFP absorption spectra \cite{Gorokhovatsky2004_BRET}---a behavior that likely extends to JFP. This parallels Venus, in which the anionic chromophore dominates at near-neutral pH \cite{Rekas2002Venus}, a feature that may underlie the strong coupling observed in Venus dimers and related functionality in wild-type proteins, and that may be captured by charge-based models \cite{Scholes2001_TrESP}. The 477--509~nm anionic absorption--emission pathway supports efficient aequorin$\to$GFP energy transfer. AvicFP1, with $\lambda_{\rm abs}\!\approx\!481~\mathrm{nm}$, may provide even better spectral overlap with aequorin than avGFP’s partially occupied 477~nm band; however, AvicFP1 is not co-expressed with aequorin \cite{Lambert2020_Aequorea}. Notably, AvicFP1 is also the closest known paralog to both GFP and JFP (formerly AausGFP), further setting it apart from other homologous FPs and CPs.

Notably, the relatively high ground-state stability of the neutral chromophore may explain why it remains accessible in both GFP and JFP and why AvicFP1 lacks obvious orthologs, aside from the potential explanation of extraneous or maladaptive regulatory mutations preventing its co-expression with aequorin \cite{Lambert2020_Aequorea,Shaner2020_ImagingData,Grigorenko2013_NeutralChromoStability}. The neutral chromophore may rather act as a “dormant stabilizer,” increasing chromophore longevity in the ground state while leaving the anionic form accessible for efficient energy transfer. The requirement for an unstable anionic state alongside a seemingly unused neutral state may also reflect historical contingencies or evolutionary anachronisms. Indeed, the $\sim$400~nm neutral absorption peak in both GFP and JFP may derive from an earlier evolutionary role as a UV/blue-light-protective dye \cite{Lambert2020_Aequorea}, with aequorin’s emission spectrum later becoming the dominant constraint on chromophore state access \cite{Gorokhovatsky2004_BRET}. 

These considerations suggest that GFP and its orthologs may have evolved as modifications of a 
pre-existing signaling pathway. Green emission underwater offers high signaling efficacy, and 
modular dimerization enhances fluorescence stability through bright-state formation. A parallel 
example exists in the more distantly related cnidarian \textit{Renilla reniformis}, which possesses its own “RrGFP,” another structurally homologous green fluorescent protein, with a single absorption peak, rather than the characteristic double peak found in \emph{Aequorea} spp., accepts energy from luciferase via BRET, serving as the primary donor, rather than aequorin \cite{Loening2007Renilla,Cubitt1995GFP}. 

It is thus rational to conclude that realistic aequorin--GFP fusion models and experiments are essential to rigorously test the adaptive and functional hypotheses presented here. Such studies may explain discrepancies between isolated GFP dimer experiments and native behavior, and could apply equally to JFP. For instance, GFP may prove to exist as an H-like dimer in the absence of aequorin, favoring dark states despite its chromophore orientation resembling Venus (Figure~\ref{fig:1MYW_1GFL_aligned}). Such a finding would deepen our understanding of the validity of the PDA in the FP distance regime, especially in the presence of a third partner such as aequorin, whose role may extend beyond electrostatic screening. The strong evidence for aequorin’s spectral influence nonetheless supports GFP and JFP as adaptive ortholog candidates, extended to Venus’s geometry and dominantly anionic absorption peak at 515~nm \cite{Rekas2002Venus,Nagai2002Venus}. Just as introducing membrane dynamics into LHC models has been shown to increase delocalization of dark states \cite{Kulkarni2025LH2}, the presence of aequorin may enhance bright-state or site delocalization, providing a further example of adaptive superradiance in nature 
\cite{Babcock2024Ultraviolet,Kalra2023AllLit}.

\textbf{Tetramers and higher-order coupling.} Additional reef coral and cephalochordate 
homologs---including DsRed from \textit{Discosoma} sp.\ and LanYFP from the ancestral cephalochordate genus, \emph{Branchiostoma} spp.---form obligate tetramers or higher-order aggregates. DsRed, for example, exhibits AB and circular dichroism spectra analogous to Venus dimers, both consistent with negative 
coupling \cite{Kim2019_VenusDimers,SanchezMosteiro2004_DsRedAntibunch,Visser2002_FEBS,Shaner2013_mNeonGreen}. 

DsRed protein structural analyses further reveal orthogonal 222 symmetry \cite{Wall2000_DsRedStruct}. The negative $J$ observed in both Venus and DsRed suggests that structural symmetry itself can enforce bright-state selection through thermal relaxation. Yet in higher-order assemblies, related symmetries can redistribute oscillator strength across multiple effective chromophore pairs. Such structural symmetries link to those present in other, more diverse groups of CPs, such as LHCs, where effectively dimeric subunits couple strongly, but higher-order $C_3$ or $C_9$ symmetries stabilize the aggregate \cite{Wendling2002_FMO,Sundstrom1999_LH2Review} while protecting dark states and energy transfer through further coupling or redundancy.

From a design perspective, tetramerization and tandem-linker engineering could stabilize dipole geometries, extend coherence time through redundant pathways, or bias exciton manifolds toward either bright or dark states depending on the desired function. This interplay between structural symmetry, excitonic entropy, and functional outcome suggests a broader organizing principle, of similar nature to those proposed previously by Hulega and Plenio \cite{Huelga2013}---in this case, connecting oligomeric structure and coupling to the management of optical information.

\textbf{Entropy, information, and J- vs.\ H-aggregates.} As currently understood, exciton aggregates, including all CPs, from FPs to LHCs, fit into either one of two broad categories, labeled J and H. J-aggregates concentrate oscillator strength into the bright state, and thermal relaxation reinforces this bias \cite{Spano2017_AggregatesBeyondKasha}. In contrast, H-aggregates distribute oscillator strength between bright and dark states, and thermal relaxation tends to funnel population into dark states that couple only weakly to light \cite{Spano2017_AggregatesBeyondKasha,Hestand2018_HJAggregates,Rouse2019_DarkStates}.

As a thermodynamic motif, the J--H distinction maps onto information entropy. J-like systems concentrate oscillator strength into a highly emissive bright-state channel, so optical measurements yield predictable outcomes (low entropy, signal-transducing). By contrast, H-like systems disperse population into dark states, leading to suppressed emission and greater uncertainty in measurement outcomes (high entropy, heat-exchanging). In this way, CP systems can either protect excited states for internal transfer or transmit them externally as optical signals.

\textbf{The bioexciton motif.} Taken together, these observations suggest a unifying principle in which biologically-adaptive CPs exploit excitonic coupling as a design rule for function---a \emph{bioexciton motif}. LHCs, for example, form dense, H-aggregate-like chains of weakly emitting states \cite{Kulkarni2025LH2,Creatore2013,Spano2017_AggregatesBeyondKasha,Sohoni2024_HAggregateBiexcitons} that suppress radiative loss and act as ``heat exchangers'' for efficient energy transport. In contrast, FPs, including Venus and DsRed, form J-like dimers \cite{Kim2019_VenusDimers,Visser2002_FEBS} which concentrate oscillator strength for optical signal transduction---a prediction extended here to other wild-type FPs, as they evolved to exist in nature. Antiparallel negative coupling, however, would flip the energies of expected J-dimers, behaving potentially more H-like in nature, although modeled presently with the energy ordering of a J-dimer \cite{Banerjee2025_DimerCoherence}. In the occurrence of a mixed state of bright and dark excitons, however, strong coupling in FPs would still lead to multiple fluorescent emission pathways, suggesting a slightly different thermodynamic function of excitonic coupling in FPs.

Exciton delocalization in LHCs is firmly supported by absorption and circular dichroism spectroscopy 
\cite{Kulkarni2025LH2,Krikunova2002_BBA,Lambert2019_RevModPhys}, and dark-state protection has both experimental and theoretical backing, including the role of delocalization in efficient energy transfer \cite{Kulkarni2025LH2,Kasha1963_ExcitonModel,Freed2025_PermDipoles,Hestand2018_HJAggregates,Rouse2019_DarkStates,Sohoni2024_HAggregateBiexcitons,Creatore2013}, though long-lived electronic coherence at physiological temperature remains debated \cite{Kim2021_QBioReview}. Viewed through this lens, the base function of a CP system can be classified by its excitonic coupling: H-like CPs favor transport, J-like CPs favor signaling, and under certain conditions J-like coupling may further optimize fluorescence quantum yield through superradiance, as demonstrated in synthetic J-aggregate nanodots with near-unity emission efficiency \cite{Piwonski2021_JAggregateJdots}. This duality links entropy flow (thermodynamics), signal-to-noise interaction (information theory), and adaptive function (evolutionary biology), establishing the bioexciton motif as both a unifying mechanistic principle and a conceptual bridge from Venus dynamics to FP-dimer-based molecular and photonic qubits.

\subsection{Technological Implications for Quantum Computing}
\label{sec:tech}

Recent work has already realized long-lived qubits in FPs by exploiting the metastable triplet state of enhanced yellow fluorescent protein (EYFP), another yellow GFP derivative. These qubit systems are capable of achieving $\mu$s coherence times and use optically detected magnetic resonance (ODMR) with optically activated delayed-fluorescence (OADF) readout, rather than standard fluorescence detection \cite{Feder2025_FluorescentProteinQubit}. In contrast, Venus-like dimers, as described here, operate in the singlet excitonic manifold, with fs--ps dynamics, well matched to ultrafast photonic gates. These two regimes---long-lived spin coherence and ultrafast excitonic coherence---are complementary, together spanning biology's accessible quantum spectrum in time by many orders of magnitude.

For an FP dimer with $|J|=34$~meV, the coherent beat oscillation period is set by the exciton splitting:
\[
T_{\rm osc} \approx \frac{2\pi\hbar}{2|J|} \sim 60~\text{fs}.
\]
Scaling to multi-qubit systems requires that coherence half-lives $t_{1/2}$ comfortably exceed gate times $T_{\rm gate}$ by at least an order of magnitude \cite{Preskill1998,NielsenChuang2010}, so that errors do not accumulate catastrophically as qubit counts grow. Thus, any gate that leverages excitonic phase (i.e. quantum coherence) must operate on 10--100~fs timescales. At 293~K, the dephasing model used throughout predicts coherence decay within a single beat cycle (within one oscillation period). Lowering $T$ reduces the Drude--Lorentz pure-dephasing rate $\gamma_\phi$ linearly in this fast-bath limit, suggesting a regime where many beat periods could be sustained and deterministic gates may become plausible.

\begin{figure}[h!tbp]
  \centering
  \includegraphics[width=1\linewidth]{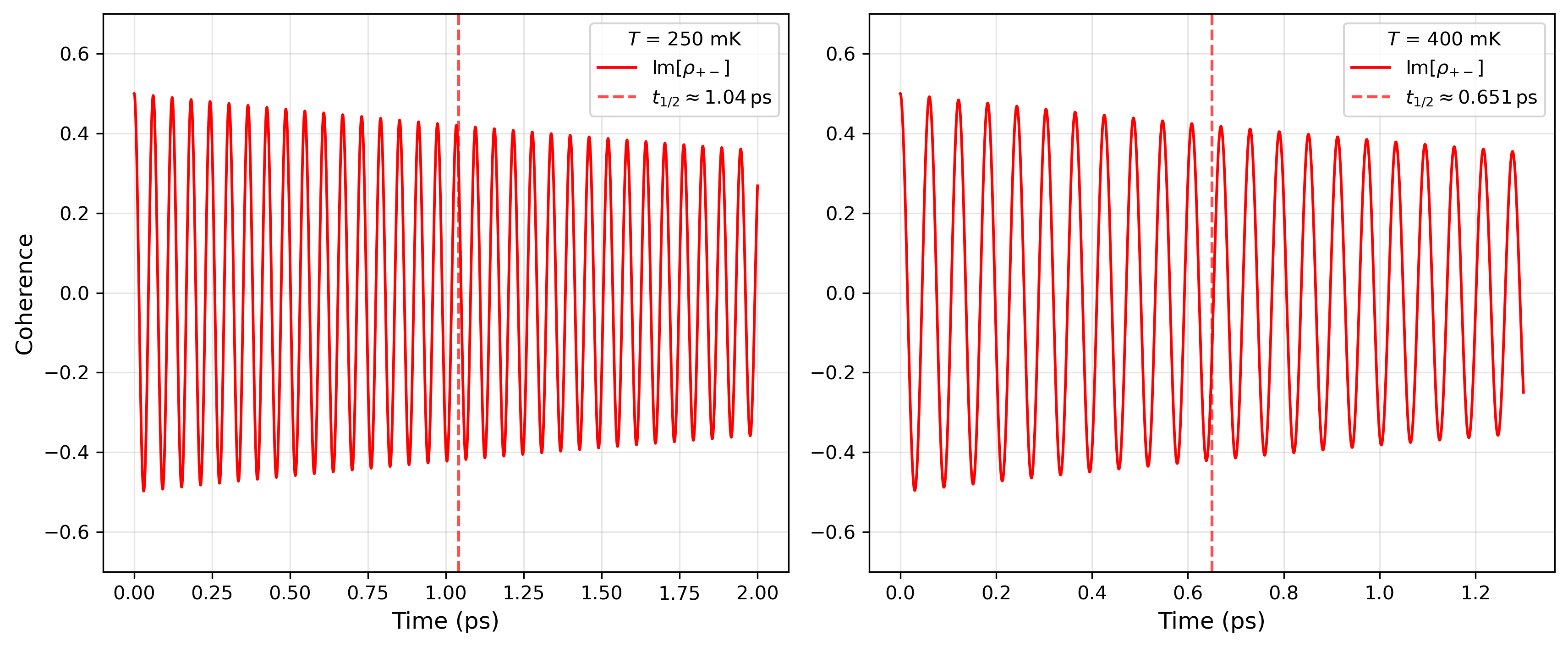}
  \caption{\textbf{Heuristic cryogenic-temperature extrapolation of coherence half-life.} Imaginary coherence Im[$\rho_{+-}$] is shown as a heuristic to highlight oscillatory decay and half-life, rather than to represent a “real” cryogenic observable in Venus dimers, both mathematically and physically. The system is initialized in the state $(|+\rangle+i|-\rangle)/\sqrt{2}$ to visualize coherence evolution. The case modeled here has identical system parameters to those used in Figure \ref{fig:room-temperature_subps_dynamics}. The high-temperature dephasing model used throughout this work is heuristically extrapolated to cryogenic temperatures (250--400 mK). Although this violates the $k_BT \gg \hbar\omega_0$ high-temperature modeling assumption, it provides a \emph{conservative lower bound:} true coherence half-lives $t_{1/2}$ would be longer than shown, since vibrations are exponentially suppressed at low $T$. Even so, the extrapolation predicts $t_{1/2}\approx 1.04$~ps at 250~mK and $t_{1/2}\approx 0.651$~ps ($651$ fs) at 400~mK, placing FP dimers within a regime well-compatible with ultrafast optical gate times.}
  \label{fig:cryogenic_exciton_dynamics}
\end{figure}

As shown in Figure~\ref{fig:cryogenic_exciton_dynamics}, the extrapolated half-lives already exceed the tens-of-fs coherence half-life by at least an order of magnitude, indicating that cryogenic operation could potentially enable \emph{many} coherent oscillations per gate. State-of-the-art cryogenic platforms already achieve electron temperatures in the 3--250~mK range on-chip via nuclear demagnetization \cite{Yurttagul2019_PRAppl}, providing ample headroom to suppress dephasing if other material parameters remain favorable. While proteins are not nanoelectronic metals, this benchmark shows that the \emph{temperature} regime needed to extend coherence time by orders of magnitude is technically reachable. This invites exploration of hybrid photonic-biomolecular experiments that test FP dimers as near-room-temperature or cryogenic qubits with ultrafast optical control.

\textbf{Current quantum computing context.} State-of-the-art photonic quantum computers already rely on fs--ps pulsed lasers to define gate operations, with characteristic timescales well matched to excitonic beating in Venus-like dimers \cite{OBrien2009_OpticalQC,Flamini2018_PhotonicReview}. Scaling to multi-qubit systems requires that coherence half-lives $t_{1/2}$ comfortably exceed gate times $T_{\rm gate}$ by at least an order of magnitude, so that errors do not accumulate catastrophically as qubit counts grow \cite{Preskill1998}. For FP-based qubits, this implies engineering conditions (temperature, geometry, environment) where $t_{1/2}/T_{\rm gate} \gtrsim 10$.

\textbf{Photonic vs.\ molecular qubits.} Photonic qubits store information in polarization, frequency, or temporal modes, while molecular qubits rely on coherent superpositions of site or exciton states \cite{NielsenChuang2010}. FP dimers sit naturally at the interface: dimerization defines a site/energy basis where excitonic coherence evolves, and emission occurs as polarized photons that can directly seed photonic quantum channels. Thus, Venus dimers provide a conceptual bridge between photonic and molecular approaches, with polarization serving as a direct experimental handle on state preparation and readout \cite{Kim2019_VenusDimers,Jimenez2023Ultrafast,Savikhin1997}.

\textbf{State preparation and design strategies.} Coherent state preparation could be achieved by ultrafast resonant pumping of the bright state \cite{Engel2007,Kim2021_QBioReview,Savikhin1997}, or by genetically engineering dimers biased toward the dark state such that emission is suppressed and coherence evolves ``silently'' until triggered. Two identical chromophores can be forced to couple with specific relative structural geometries to yield red- and blueshifted excitons (bright and dark states, respectively). Genetic variants that stabilize dark-favored manifolds may thus allow extended coherent evolution by evading radiative decay.

\textbf{Forward-looking implications.} The present analysis heuristically probes the cryogenic temperature limit and highlights timescales on which quantum gate operations appear feasible. Moving forward, further theoretical and practical exploration of regimes where $t_{1/2}$ can be extended---through temperature control, engineered rigidity, or targeted chromophore environments---will be essential for assessing whether FP dimers can achieve the $t_{1/2}/T_{\rm gate}$ ratios needed for fault-tolerant operation. In principle, the same structural properties that evolution exploited for spectral tuning (chromophore geometry, oligomerization, etc.) may be repurposed as design principles for molecular qubits operating at the interface with photonic quantum technologies in addition to spin. Hence, the same design tricks \textit{Aequorea} jellyfish use to shine green in the ocean may one day be harnessed to sustain coherence in quantum computers.

\end{document}